% FM 00-2; mp_arc \# 00-60; nlin/CD \# 0001???
%**start of header
%\input fiat
%**end of header

\newcount\mgnf\newcount\tipi\newcount\tipoformule
\newcount\aux\newcount\piepagina\newcount\xdata
\mgnf=0
\aux=1           %1 produce aux
\tipoformule=0   %0 usa aux; 1 no (usa i simboli dati)
\piepagina=1     %0 =data e #par.#pag; 1=data e #pag; 2=#pag
\xdata=0         %0 data del giorno, 1 data fissa da \Di:
\def\Di{}

\ifnum\mgnf=1 \aux=0 \tipoformule =1 \piepagina=1 \xdata=1\fi
\newcount\bibl
%\bibl= ?               % 0= rif [XXX], 1= rif. numerici
\ifnum\mgnf=0\bibl=0\else\bibl=1\fi
\bibl=0

% Per poter cambiare a piacimento il formato dei riferimenti
% bibliografici in <nome>.tex:
%
% 1: citare nella forma esemplificata da \ref{B}{2}{20}}
%    ove XXX e' un simbolo per le iniziali e 2 distingue i lavori con
%    le stesse iniziali,  7 e' il numero SIMBOLICO del riferimento per XXX2.
%    Il numero 7 puo' essere ARBITRARIO e viene automaticamente
%    riaggiustato al momento della compilazione (vedi punto 4)
%
% 2: Se si sceglie \bibl=0 si cita nella forma [XXX2]; se si sceglie
%    \bibl=1 si cita nella forma [numero di ordine di prima citazione].
%
% 3: La bibliografia va scritta nella forma \def{\qqq}{<citazione>}
%    in ordine alfabetico per autore attribuendo un simbolo
%    qualsiasi al testo <citazione} (ad es \def\REFCIT{citazione} e
%    va seguita dal comando  \rif{XXX}{2}{\qqq}{0}
%    scritto all' inizio di una riga altrimenti bianca. L' ultimo {0}
%    e' un parametro inutile ed e' li solo provvisoriamente per memoria
%    (di tentativi in corso).
%
% 4: Per il riaggiustamento automatico (in ordine di citazione)
%    occorre compilare (ottenendo oltre ai soliti la scheda ausiliaria
%    ref.b) e poi si deve eseguire il
%    programma rif <nome> che (usando ref.b) produce fin.tex e <nome>.tex
%    con i riferimenti giusti
%    in \bibl=1 e la si ricompila e stampa. La scheda iniziale <nome>.tex
%    diventa <nome>.old.
\ifnum\bibl=0
\def\ref#1#2#3{[#1#2]\write8{#1@#2}}
\def\rif#1#2#3#4{\item{[#1#2]} #3}
\fi

\ifnum\bibl=1
\openout8=ref.b
\def\ref#1#2#3{[#3]\write8{#1@#2}}
\def\rif#1#2#3#4{}

\fi

\def\9#1{\ifnum\aux=1#1\else\relax\fi}
\ifnum\piepagina=0 \footline={\rlap{\hbox{\copy200}\
$\st[\number\pageno]$}\hss\tenrm \foglio\hss}\fi \ifnum\piepagina=1
\footline={\rlap{\hbox{\copy200}} \hss\tenrm \folio\hss}\fi
\ifnum\piepagina=2\footline{\hss\tenrm\folio\hss}\fi

\ifnum\mgnf=0 \magnification=\magstep0
\hsize=13.5truecm\vsize=22.5truecm \parindent=4.pt\fi
\ifnum\mgnf=1 \magnification=\magstep1
\hsize=16.0truecm\vsize=22.5truecm\baselineskip14pt\vglue5.0truecm
\overfullrule=0pt \parindent=4.pt\fi

\let\a=\alpha\let\b=\beta \let\g=\gamma \let\d=\delta
\let\e=\varepsilon \let\z=\zeta 
\let\th=\vartheta\let\k=\kappa \let\l=\lambda \let\m=\mu \let\n=\nu
 \let\p=\pi  \let\s=\sigma \let\t=\tau
 \let\f=\varphi\let\ch=\chi \let\ps=\psi \let\o=\omega
 \let\G=\Gamma \let\D=\Delta 
\let\L=\Lambda   
  
{\count255=\time\divide\count255 by 60 \xdef\oramin{\number\count255}
\multiply\count255 by-60\advance\count255 by\time
\xdef\oramin{\oramin:\ifnum\count255<10 0\fi\the\count255}}
\def\ora{\oramin }

%\Di e' definito all' inizio e ridefinito qui
\ifnum\xdata=0
\def\data{\number\day/\ifcase\month\or gennaio \or
febbraio \or marzo \or aprile \or maggio \or giugno \or luglio \or
agosto \or settembre \or ottobre \or novembre \or dicembre
\fi/\number\year;\ \ora}
\def\Di{\number\day\kern2mm\ifcase\month\or gennaio \or
febbraio \or marzo \or aprile \or maggio \or giugno \or luglio \or
agosto \or settembre \or ottobre \or novembre \or dicembre
\fi\kern0.1mm\number\year}
\else
\def\data{\Di}
\fi

\setbox200\hbox{$\scriptscriptstyle \data $}
\newcount\pgn \pgn=1
\def\foglio{\number\numsec:\number\pgn
\global\advance\pgn by 1} \def\foglioa{A\number\numsec:\number\pgn
\global\advance\pgn by 1}
\global\newcount\numsec\global\newcount\numfor \global\newcount\numfig
\gdef\profonditastruttura{\dp\strutbox}
\def\senondefinito#1{\expandafter\ifx\csname#1\endcsname\relax}
\def\SIA #1,#2,#3 {\senondefinito{#1#2} \expandafter\xdef\csname
#1#2\endcsname{#3} \else \write16{???? ma #1,#2 e' gia' stato definito
!!!!} \fi} \def\etichetta(#1){(\veroparagrafo.\veraformula) \SIA
e,#1,(\veroparagrafo.\veraformula) \global\advance\numfor by 1
\9{\write15{\string\FU (#1){\equ(#1)}}} \9{ \write16{ EQ \equ(#1) == #1
}}} \def \FU(#1)#2{\SIA fu,#1,#2 }
\def\etichettaa(#1){(A\veroparagrafo.\veraformula) \SIA
e,#1,(A\veroparagrafo.\veraformula) \global\advance\numfor by 1
\9{\write15{\string\FU (#1){\equ(#1)}}} \9{ \write16{ EQ \equ(#1) == #1
}}} \def\getichetta(#1){Fig.  \verafigura \SIA e,#1,{\verafigura}
\global\advance\numfig by 1 \9{\write15{\string\FU (#1){\equ(#1)}}} \9{
\write16{ Fig.  \equ(#1) ha simbolo #1 }}} \newdimen\gwidth \def\BOZZA{
\def\alato(##1){ {\vtop to \profonditastruttura{\baselineskip
\profonditastruttura\vss
\rlap{\kern-\hsize\kern-1.2truecm{$\scriptstyle##1$}}}}}
\def\galato(##1){ \gwidth=\hsize \divide\gwidth by 2 {\vtop to
\profonditastruttura{\baselineskip \profonditastruttura\vss
\rlap{\kern-\gwidth\kern-1.2truecm{$\scriptstyle##1$}}}}} }
\def\alato(#1){} \def\galato(#1){}
\def\veroparagrafo{\number\numsec}\def\veraformula{\number\numfor}
\def\verafigura{\number\numfig}
\def\geq(#1){\getichetta(#1)\galato(#1)}
\def\Eq(#1){\eqno{\etichetta(#1)\alato(#1)}}
\def\eq(#1){\etichetta(#1)\alato(#1)}
\def\Eqa(#1){\eqno{\etichettaa(#1)\alato(#1)}}
\def\eqa(#1){\etichettaa(#1)\alato(#1)}
\def\eqv(#1){\senondefinito{fu#1}$\clubsuit$#1\write16{No translation
for #1} \else\csname fu#1\endcsname\fi}
\def\equ(#1){\senondefinito{e#1}\eqv(#1)\else\csname e#1\endcsname\fi}
\openin13=#1.aux \ifeof13 \relax \else \input #1.aux \closein13\fi
\openin14=\jobname.aux \ifeof14 \relax \else \input \jobname.aux
\closein14 \fi \9{\openout15=\jobname.aux} \newskip\ttglue

%\font\dodicirm=cmr12\font\dodicibf=cmbx12\font\dodiciit=cmti12
%\font\titolo=cmbx12 scaled \magstep2
%\font\ottorm=cmr8\font\ottoi=cmmi8\font\ottosy=cmsy8
%\font\ottobf=cmbx8\font\ottott=cmtt8\font\ottosl=cmsl8\font\ottoit=cmti8
%\font\sixrm=cmr6\font\sixbf=cmbx6\font\sixi=cmmi6\font\sixsy=cmsy6

\font\titolo=cmbx10 scaled \magstep1
\font\ottorm=cmr8\font\ottoi=cmmi7\font\ottosy=cmsy7
\font\ottobf=cmbx7\font\ottott=cmtt8\font\ottosl=cmsl8\font\ottoit=cmti7
\font\sixrm=cmr6\font\sixbf=cmbx7\font\sixi=cmmi7\font\sixsy=cmsy7

\font\fiverm=cmr5\font\fivesy=cmsy5\font\fivei=cmmi5\font\fivebf=cmbx5
\def\ottopunti{\def\rm{\fam0\ottorm}\textfont0=\ottorm%
\scriptfont0=\sixrm\scriptscriptfont0=\fiverm\textfont1=\ottoi%
\scriptfont1=\sixi\scriptscriptfont1=\fivei\textfont2=\ottosy%
\scriptfont2=\sixsy\scriptscriptfont2=\fivesy\textfont3=\tenex%
\scriptfont3=\tenex\scriptscriptfont3=\tenex\textfont\itfam=\ottoit%
\def\it{\fam\itfam\ottoit}\textfont\slfam=\ottosl%
\def\sl{\fam\slfam\ottosl}\textfont\ttfam=\ottott%
\def\tt{\fam\ttfam\ottott}\textfont\bffam=\ottobf%
\scriptfont\bffam=\sixbf\scriptscriptfont\bffam=\fivebf%
\def\bf{\fam\bffam\ottobf}\tt\ttglue=.5em plus.25em minus.15em%
\setbox\strutbox=\hbox{\vrule height7pt depth2pt width0pt}%
\normalbaselineskip=9pt\let\sc=\sixrm\normalbaselines\rm}

\catcode`@=11
\def\footnote#1{\edef\@sf{\spacefactor\the\spacefactor}#1\@sf
\insert\footins\bgroup\ottopunti\interlinepenalty100\let\par=\endgraf
\leftskip=0pt \rightskip=0pt \splittopskip=10pt plus 1pt minus 1pt
\floatingpenalty=20000
\smallskip\item{#1}\bgroup\strut\aftergroup\@foot\let\next}
\skip\footins=12pt plus 2pt minus 4pt\dimen\footins=30pc\catcode`@=12
\let\nota=\ottopunti

%% Grafica
\newdimen\xshift \newdimen\xwidth \newdimen\yshift

\def\ins#1#2#3{\vbox to0pt{\kern-#2 \hbox{\kern#1
#3}\vss}\nointerlineskip}

\def\eqfig#1#2#3#4#5{ \par\xwidth=#1
\xshift=\hsize \advance\xshift by-\xwidth \divide\xshift by 2
\yshift=#2 \divide\yshift by 2 \line{\hglue\xshift \vbox to #2{\vfil #3
\includegraphics{#4.ps} }\hfill\raise\yshift\hbox{#5}}}

\def\dsgn#1#2#3#4{%
\kern-#1\,\,\hbox to #1{ \vbox to #2{\vfil #3%
\includegraphics{#4.ps} }}\kern#1}%#1 dim orizz. #2 dim vert., #3
				 %inserti, #4 scheda

\let\propagatore=\dsgn

\def\8{\write13}

\def\didascalia#1{\vbox{\nota\0#1\hfill}\vskip0.3truecm}

%%%%%%%%%%%%%%%%%%%%%%

\def\V#1{{\,\underline#1\,}}
\def\T#1{#1\kern-4pt\lower9pt\hbox{$\widetilde{}$}\kern4pt{}}
\let\dpr=\partial\def\Dpr{{\V\dpr}} \let\io=\infty\let\ig=\int
\def\fra#1#2{{#1\over#2}}\def\media#1{\langle{#1}\rangle}\let\0=\noindent
\def\guida{\leaders\hbox to 1em{\hss.\hss}\hfill}
\def\tende#1{\vtop{\ialign{##\crcr\rightarrowfill\crcr
\noalign{\kern-1pt\nointerlineskip} \hglue3.pt${\scriptstyle
#1}$\hglue3.pt\crcr}}} \def\otto{\
{\kern-1.truept\leftarrow\kern-5.truept\to\kern-1.truept}\ }

\def\pagina{\vfill\eject}\let\ciao=\bye

\def\st{\scriptscriptstyle}
\def\*{\vskip0.3truecm}

\def\lis#1{{\overline #1}}\def\eg{\hbox{\it e.g.\ }}
\def\ap{\hbox{\it a priori\ }}
\def\ie{\hbox{\it i.e.\ }}

\def\fiat{{}}
\def\\{\hfill\break} \def\={{ \; \equiv \; }}

\def\annota#1{\footnote{${{}^{\bf#1}}$}}
\ifnum\aux=1\BOZZA\else\relax\fi
\ifnum\tipoformule=1\let\Eq=\eqno\def\eq{}\let\Eqa=\eqno\def\eqa{}
\def\equ{{}}\fi
\def\defi{\,{\buildrel def \over =}\,}

\def\1{\ifnum\mgnf=0\pagina\else\relax\fi}
\def\W#1{#1_{\kern-3pt\lower6.6truept\hbox to 1.1truemm
{$\widetilde{}$\hfill}}\kern2pt\,}

\def\FINE{
\*
\*
\0{\it Internet:
Author's  preprints at:  {\tt http://ipparco.roma1.infn.it}

\0\sl e-mail: {\tt giovanni.gallavotti@roma1.infn.it}}}

\def\aa{{\V \a}}\def\nn{{\V\n}}\def\AA{{\V A}}
\def\pps{{\V \ps}}
\def\oo{{\V \o}}\def\xx{{\V x}}
\def\HH{{\cal H}}\def\CC{{\cal C}}\def\LL{{\cal L}}\def\BB{{\cal B}}
\def\kk{{\V k}}

\def\dsgn#1#2#3#4{\xshift=#1%
\yshift=#2 \divide\yshift by 2 \hbox{ \vbox to #2{\vfil #3%
\includegraphics{#4.ps} }}\kern\xshift}

\let\propagatore=\dsgn
\def\cfr{{\it c.f.r. }}
\fiat

\vglue1truecm

\centerline{\titolo Renormalization group in Statistical Mechanics and}
\centerline{\titolo Mechanics: gauge symmetries and vanishing beta
functions.}
\*\*
\centerline{\it Giovanni Gallavotti}
\*
\centerline{Fisica, Universit\`a di Roma 1}
\centerline{P.le Moro 2, 00185 Roma, Italia}
\*\*
\0{\bf Abstract:\it
Two very different problems that can be studied by renormalization group
methods are discussed with the aim of showing the conceptual unity that
renormalization group has introduced in some areas of theoretical
Physics. The two problems are: the ground state theory of a one
dimensional quantum Fermi liquid and the existence of quasi periodic
motions in classical mechanical systems close to integrable ones. I
summarize here the main ideas and show that the two treatments, although
completely independent of each other, are strikingly similar.}
\*\*

\0{\bf\S1. Introduction.}
\numsec=1\numfor=1\*

There are few cases in which a renormalization group analysis can be
performed in full detail and without approximations.  The best known
case is the {\it hierarchical model} theory of Wilson, [Wi70], [WK74].
Other examples are the (Euclidean) $\f^4$ quantum field theories in two
and three space--time dimensions, [Wi73] (for an analysis in the
spirit of what follows see [Ga85] or [BG95]), and the universality
of critical points [WF72].  In all such examples there is a basic
difficulty to overcome: namely the samples of the fields can be
unboundedly large: this does not destroy the method because such large
values have extremely small probability, [Ga85].  The necessity
of a different treatment of the large and the small field values hides,
to some extent, the intrinsic simplicity and elegance of the approach:
unnecessarily so as the end result is that one can essentially ignore
(to the extent that it is not even mentioned in most application
oriented discussions) the large field values and treat the
renormalization problem perturbatively, as if the large fields were not
possible.
\*

Here I shall discuss two (non trivial) problems in which the large
field difficulties are not at all present, and the theory leads to a
convergent perturbative solution of the problem (unlike the the above
mentioned classical cases, in which the perturbation expansion cannot
be analytic in the perturbation parameter).

The problems are:
\*

\0(1) the theory of the ground state of a system of (spinless, for
simplicity) fermions in $1$-dimension, [BGPS],[BG90],[BM95];
\\
(2) the theory of KAM tori in classical mechanics, [El96], [Ga95],
[BGGM], [GGM95].

\*
The two problems will be treated {\it independently}, for
completeness, although it will appear that they are closely related.
Since the discussion of problem (1) is quite technical we summarize it
at the end (in \S3) in a form that shows the generality of the method
that will then be applied to the problem (2) in \S4.

The analysis of the above examples suggests methods to study and solve
several problems in the theory of rapidly
perturbed quasi periodic unstable motion, [Ga95], [GGM99]: but for
brevity we shall only refer to the literature for such applications.
\*
\1

\0{\bf\S2. Fermi systems in one dimension.}
\numsec=2\numfor=1
\*

The Hamiltonian for a system of $N$ {\it spinless} fermions at
$\xx_1,\ldots,\xx_N$ enclosed in a box (actually an interval) of size
$V$ is:
$$H=\sum_{i=1}^N \big(\fra1{2m}\D_{\xx_i}-\m\big)+ 2\l\sum_{i<j}
v(\xx_i-\xx_j)- \sum_{i=1}^N \n\Eq(2.1)$$
where $\m$ is the chemical potential, $v$ is a smooth interaction pair
potential, $\l$ is the strength of the coupling; $\n$ is a correction
to the chemical potential that vanishes for $\l=0$ and that has to be
adjusted as a function of $\l$; it is introduced in order that
the Fermi momentum stays $\l$--independent and equal, therefore, to
$p_F=(2m \m)^{\fra12}$. It is in fact convenient to develop the theory
at fixed Fermi momentum because the latter has a more direct physical
meaning than the chemical potential as it marks the location of
important singukarities of the functions that describe the theory. The
parameter $m$ is the mass of the particles in absence of interaction.

It is well known, [LW60], that the ground state of the above Hamiltonian
is described by the Schwinger functions of a fermionic theory
whose fields will be denoted $\ps^\pm_x$.
For instance the {\it occupation number} function $n_\kk$ which, in
absence of interaction, is the simple characteristic function
$n_\kk=1$ if $|\kk|<p_F$ and $n_\kk=0$ if $|\kk|>p_F$ is, in general,
the Fourier transform of $S(\xx,t)$ with
$x=(\xx,t)=(\xx_1-\xx_2,t_1-t_2)$ and $t=0^+$:
$$S(\xx)=S(\xx_1,t_1;\xx_2,t_2)\big|_{t_1=t_2^+}=
\lim_{\b\to\io\atop V\to\io}
\fra{{\rm Tr}\, e^{-(\b -t_1)H}\ps^+_{x_1,t_1}
e^{-(t_1-t_2) H} \ps^-_{x_2,t_2} e^{-t_2 H}}
{{\rm Tr}\, e^{-\b H}}\Big|_{t_1=t_2^+}\Eq(2.2)$$
Formal perturbation analysis of the $2$--points Schwinger function
$S(x)$ and of the $n$--points natural extensions
$S(x_1,x_2,\ldots,x_n)$ can be done and the (heuristic) theory is very
simple in terms of Feynman graphs.

The $n$-points Schwinger function is expressed as a power series in
the couplings $\l,\n$ $\sum_{p=0}^\io \l^p\n^q\,
S^{(p,q)}(x_1,x_2,\ldots,x_n)$ with the coefficients $S^{(p,q)}$
computed by considering the (connected) Feynman graphs composed by
linking together in all possible ways the following basic ``{\it graph
elements}''
\*

(1) $p$ ``internal $4$--lines graph elements'' (also called ``coupling
graphs'') and $q$ ``internal $2$--lines graph elements'' (or ``chemical
potential vertices'') of the form:

\*
\eqfig{230pt}{60pt}{
\ins{15pt}{10pt}{$x$}
\ins{195pt}{10pt}{$x$}
\ins{18pt}{58pt}{$x'$}
}{nodi}{}
\*
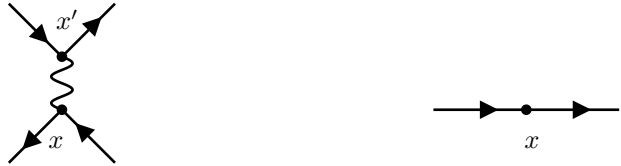
\0Fig. 1: {\nota The two basic building blocks (``graph elements'') of
the the Feynman graphs for the description of the ground state: the
first represents the potential term ($2\l v$) in \equ(2.1)) and the
second the chemical potential term ($\n$).}
\*

\0where the incoming or outgoing arrows represent $\ps^-_x$ or
$\ps^+_x$, respectively, and

(2) $n$ single lines attached to ``external'' vertices $\xx_j$:
the first half of which oriented towards the vertex $x$ and the other
half of them oriented away from it:

\eqfig{130pt}{0pt}{
\ins{40pt}{-5pt}{$x$}
\ins{100pt}{-5pt}{$x$}
}{ext}{}
\*
Fig. 2: {\nota Graphical representation of the ``external'' lines and
vertices in Feynman graphs.}
\*

\0The graphs are formed by {\it contracting} (\ie joining) together
lines with equal orientation. The lines emerging from different nodes
are regarded as distinct: we can imagine that each line carries a
label distinguishing it from any other, \eg the lines are thought to
be numbered from $1$ to $4$ or from $1$ to $2$, depending on the
structure of the graph element to which they belong. So that there are
many graphs giving the same contributions.

Each graph is assigned a {\it value} which is $\pm
(p!q!)^{-1}\l^p\n^q$ times a product of {\it propagators}, one per
line.  The propagator for a line joining $x_1$ to $x_2$ is, if
$x_1=(\xx_1,t_1), \, x_2=(\xx_2,t_2)$:
$$g(x_1-x_2)=
{}_{x_1}{\propagatore{40pt}{5pt}{}{x}}_{\hskip5pt x_2}
=\fra1{(2\p)^2}\ig\fra{
e^{-i(k_0 (t_1-t_2)+\kk(\xx_1-\xx_2))}}{-i
k_0+(\kk^2-p_F^2)/{2m}}\,dk_0d\kk\Eq(2.3)$$
A {\it wavy line}, see Fig. 1, joining $x_1$ with $x_2$ is
also given a propagator

$$\tilde g(x_2-x_1)= v(\xx_2-\xx_1)\,\d(t_2-t_1)\Eq(2.4)$$
associated with the ``potential''. However the wavy lines are necessarily
internal as the can only arise from the first graph element in
Fig. 1.

The $p+q$ {\it internal} node labels $(\V x,t)$ must be integrated
over the volume occupied by the system (\ie the whole space--time when
$V,\b\to\io$): the result will be called the ``integrated value'' of
the graph or simply, if not ambiguous, the graph value.

Since the value of a graph has to be integrated over the labels
$x=(\xx,t)$ of the internal nodes we shall often consider also the
value of a graph $\th$ {\it without the propagators corresponding to
the external lines} but integrated with respect to the positions of
all nodes that are not attached to external lines and we call it the
``{\it kernel}'' of the graph $\th$: the value of a graph $\th$ will
often be denoted as ${\rm Val}\,\th$ and the kernel by $K_\th$. Note
that the kernel of a graph depends on less variables: in particular it
depends only on the positions of the internal nodes; it also depends
on the labels $\oo$ of the branches external to them through which
they are connected to the external vertices.

Introducing the notion of kernel is useful because it makes natural to
collect together values of graphs which contain subgraphs
with the same number of lines exiting them, \ie whose kernels have the
same number of variables.
\*

The function $\l^p\n^q S^{(p,q)}$ is given by the sum of the
values of all Feynman graphs with $p$ vertices of the first type in
Fig. 1 and $q$ of the second type in Fig. 1 and, of course, $n$
external vertices, {\it integrated} over the internal vertices
positions. As an example consider the following contribution
to $S^{(4,2)}$:

\eqfig{180pt}{75pt}{
\ins{8pt}{15pt}{$\st x_1$}
\ins{0 pt}{60pt}{$\st x_2$}
\ins{45pt}{21pt}{$\st x_3 $}
\ins{45pt}{50pt}{$\st x_3'$}
\ins{56pt}{33pt}{$\st x_4$}
\ins{53pt}{66pt}{$\st x_4'$}
\ins{83pt}{33pt}{$\st x_5$}
\ins{86pt}{64pt}{$\st x_5'$}
\ins{98pt}{37pt}{$\st x_6$}
\ins{118pt}{21pt}{$\st x_7$}
\ins{118pt}{52pt}{$\st x_7'$}
\ins{168pt}{8pt}{$\st x_8$}
\ins{168pt}{58pt}{$\st x_9$}
\ins{23pt}{62pt}{$\st x_{10}$}}
{esempio}{}

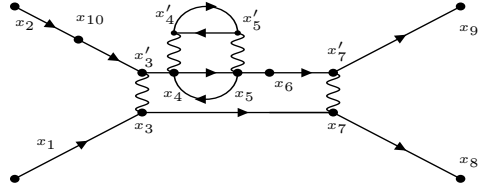
\0{\it Fig. 3: \nota An example of a Feynman graph: in spite of its
involved structure it is far simpler than its numerical expression, see
\equ(2.5). A systematic consideration of graphs as ``short cuts'' for
formulae permits us to visualize more easily various quantities and
makes it possible to recognize cancellations due to symmetries.}
\*

The value of the above graph can be easily written in formulae: apart
from a global sign that has to be computed by a careful examination of
the order in which the $x_j$-labels are written it is
$$\eqalign{
S^{(4,2)}_\th&(x_1,x_2,x_8,x_9)=\pm\fra1{4!\,2!}\ig
\,g(x_{10}-x_2)\,g(x_3-x_1)\,g(x'_3-x_{10})\cdot\cr
&\cdot
g(x_4-x'_3)\,g(x_5-x_4)\,g(x_4-x_5)\,g(x'_5-x'_4)\,g(x'_4-x'_5)\,
g(x_6-x_5)\cdot\cr &\cdot g(x_4-x_5)\,g(x_5-x_4)\,
g(x'_7-x_6)\,g(x_9-x'_7)\,g(x_8-x_7)\,\d(t_4-t'_4)\cdot\cr
&\cdot\d(t_5-t'_5)\, v(\xx_3'-\xx_3)\,\d(t_{3}'-t_3)\,
v(\xx_7'-\xx_7)\,\d(t_{7}'-t_7)\cdot\cr &\cdot
v(\xx_4'-\xx_4)\,v(\xx_5'-\xx_5)\,
dx_3\,dx'_3\, dx_4\,dx'_4\, dx_5\, dx'_5\, dx_6\, dx_7\,
dx'_7dx_{10}\cr }\Eq(2.5)$$
which is easily derived from the figure. And one hardly sees how this
formula could be useful, particularly if one thinks that this is but
{\it one} of a large number of possibilities that arise in evaluating
$S$: not to mention what we shall get when looking at higher orders, \ie
at $S^{(p,q)}$ when $p$ is a bit larger than $2$.
\*

Many (in fact most) of the integrals over the node variables $x_v$
will, however, {\it diverge}. This is a "trivial" divergence due to
the fact that interaction tends to change the value of the chemical
potential.  The chemical potential is related (or can be related) to
the Fermi field propagator singularities, and the chemical potential
is changed (or may be changed) by the interaction: the divergences
are due to the naivet\'e of the attempt at expanding the functions $S$
in a power series involving functions with singularities located ``at
the wrong places''.

The divergences disappear if the (so far free) parameter $\n$ is
chosen
to depend on $\l$ as:

$$\n=\sum_{k=1}^\io \n_k \l^k\Eq(2.6)$$
with the coefficients $\n_k$ suitably defined so that the resulting
power series in the single parameter $\l$ has coefficients free of
divergences, [LW60].
\*

This leads to a power series in just one parameter $\l$ and the "only"
problem left is therefore that of the convergence of the expansion of
the Schwinger functions in powers of $\l$. This is non trivial because
naive estimates of the sum of all graphs contributing to a given order
$p$ yield bounds that grow like $p!$, thus giving a vanishing estimate
for the radius of convergence.
\*

The idea is that there are cancellations between the values of the
various graphs contributing to a given order in the power series for the
Schwinger functions: and that such cancellations can be best exhibited
by further breaking up the values of the graphs and by again combining
them conveniently.

The "renormalization group method" can be seen in different ways: here I
am proposing to see it as a resummation method for (possibly divergent)
power series.

Keeping the original power series in $\l,\n$, \ie postponing the
choice of $\n$ as a function of $\l$, one checks the elementary fact
that the propagator $g(x)$ can be written, setting $k=(k_0,\kk)\in
R^2$, also as:
$$\eqalign{ g(x)=&\sum_{h=-\io}^1\sum_{\oo=\pm1}
e^{i\oo\,p_F\xx} {2^h} g^{(h)}_\oo(2^hp_F x)\cr
\hat g^{(h)}_\oo(k)=&\fra{\chi^{(h)}(k)}{-i k_0+\oo \kk}+{\rm ``neglegible\
corrections''} \cr} \Eq(2.7)$$
where $\chi^{(1)}(k)$ is a function increasing from $0$ to $1$ between
$\fra12 p_F$ and $p_F$, while the functions $\chi^{(h)}(k)$ are the
same function scaled to have support in $2^{h-2} p_F<|k|< 2^h p_F$.
This means that for $h\le0$ it is $\chi^{(h)}(k)=\chi(2^{-h}k
p_F^{-1})$. The simplest choice is to take $\ch^{(1)}(k)$ to be the
characteristic function of $z\=p_F^{-1}|k|>1$ and $\chi(z)$ to be the
characteristic function of the interval $[\fra12,1]$:

\eqfig{280pt}{70pt}{
\ins{28pt}{64pt}{$\ch$}
\ins{170pt}{64pt}{$\ch^{(1)}$}
\ins{36pt}{10pt}{$1/2$}
\ins{66pt}{10pt}{$1$}
\ins{200pt}{10pt}{$1$}
}
{caratter}{}

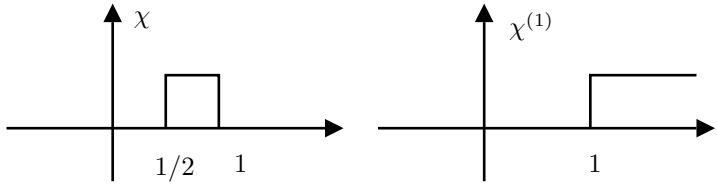
\0{Fig. 4: \nota A (non smooth)
scaling (by a factor of $2$) decomposition of unity.}

\0so that

$$\sum_{h=-\io}^1\ch^{(h)}(k)\=1\Eq(2.8)$$
To avoid technical problems it would be convenient to smoothen the
discontinuities in Fig. 4 of $\ch$ and $\ch^{(1)}$ turning them into
$C^\io$--functions which in a small vicinity of the jump increase from
$0$ to $1$ or decrease from $1$ to $0$, this is possible while {\it
still keeping the scaling decomposition} \equ(2.8) (\ie with
$\ch^{(h)}(k)\=\ch(k)$). However the formalism that this smoothing
would require is rather havy and hides the stucture of the approach;
therefore we shall continue with the decomposition of unity in
\equ(2.8) with the sharply discontinuous functions in Fig. 4, warning
(\cfr footnote ${}^2$ below) the reader when this should cause a
problem.

The ``negligible terms'' in \equ(2.7) are terms of a similar form but
which are smaller by a factor $2^h$ at least: their presence does not
alter the analysis other than notationally. {\it We shall henceforth
set them equal to $0$ because taking them into account only introduces
notational complications}.
\*

The above is an {\it infrared scale decomposition} of the propagator
$g(x)$: in fact the propagator $g^{(h)}$ contains only momenta $k$ of
$O(2^h p_F)$ for $h\le0$ while the propagator $g^{(1)}$ contains all
(and only) large momenta (\ie the {\it ultraviolet} part of the
propagator $g(x)$). The representation \equ(2.7) is called a {\it
quasi particles} representation of the propagator and the quantities
$\oo\, p_F$ are called a {\it quasi particles} momenta. The function
$g^{(h)}_\oo$ is the ``{\it quasi--particle propagator on scale $h$}''.

After extracting the exponentials $e^{i\oo p_F \xx}$ from the propagators
the Fourier transforms $\hat g^{(h)}_\oo(k)$ of $g^{(h)}_\oo(\xx)$ will no
longer be oscillating on the scale of $p_F$ and the variable $k$ will
have the interpretation of ``momentum measured from the Fermi surface''.
\*

The mentioned divergences are {\it still present} because we do not yet
relate $\l$ and $\n$: they will be eliminated temporarily by introducing
an {\it infrared cut--off}: \ie by truncating the sum in
\equ(2.7) to $h\ge -R$. We then proceed keeping in mind that {\it we must
get results which are uniform as} $R\to\io$: this will be eventually
possible only if $\n$ is suitably fixed as a function of $\l$.

Writing $g(x)={Z_1^{-1}} g^{(1)}(x)+{Z_1^{-1}}g^{(\le0)}(x)$ with
$Z_1\defi1$ and $g^{(\le m)}$ being defined in general, see \equ(2.7),
as:

$$g^{(\le m)}(x)=\sum_{h=-R}^m\sum_{\oo=\pm1} 2^h
  e^{i\oo\,p_F\xx}\,  g^{(h)}_\oo(2^hp_F x),\qquad m\le0\Eq(2.9)$$
each graph can now be decomposed as a sum of graphs each of which with
internal lines carrying extra labels ``$1$'' and ``$\oo$'' or
``$\le0$'' and ``$\oo$'' (signifying that the value of the graph has to
be computed by using the propagator ${Z_1}^{-1} g_\oo^{(1)}(x-x')$ or
${Z_1}^{-1} g_\oo^{(\le0)}(x-x')$ for the line in question, if it goes
from $x'$ to $x$).
\*

We now define {\it clusters of scale $1$}: a ``cluster'' on scale $1$
will be any set $C$ of vertices connected by lines bearing the scale
label $1$ and which are maximal in size (\ie they are not part of
larger clusters of the same type). {\it Wavy lines are regarded as
bearing a scale label $1$}. The graph is thus decomposed into smaller
graphs formed by the clusters and connected by lines of scale $\le0$:
it is convenient to visualize the clusters as enclosed into contours
that include the vertices of each cluster as well as all the lines
that connect two vertices of the same cluster.  The latter can be
naturally called {\it lines internal to the cluster} $C$.

The integrated value of a graph will be represented, up to a sign
which can be determined as described above, as a sum over the quasi
particles labels $\oo$ of the cluster lines and as an integral over the
locations of the inner vertices of the various clusters lines. The
integrand is a product between
\\
(a) the {\it kernels} $K_{C_i}$ associated with the clusters $C_i$ and
depending {\it only} on the locations of the vertices inside the
cluster $C_i$ which are extremes of lines external to the cluster and
on the quasi particles labels $\oo$ of the lines that emerge from
it,\annota{1}{\nota By definition the kernel $K_C$ also involves
integration over the locations of its inner vertices and the sum over
the quasi particle momenta of the inner propagators.}
and
\\
(b) the propagators $Z_1^{-1} g_\oo^{(\le0)}$ corresponding to the
lines that are external to the clusters (in the sense that they have
at least one vertex not indide the cluster).
\*

We now look at the clusters $C$ that have just $|C|=2$ or $|C|=4$
external lines and that are therefore associated with kernels
$K_C(\{x_j,\oo_j\}_{j=1,2};C)$ or $K_C(\{x_i,\oo_i\}_{i=1,\ldots,4};C)$.
Such kernels, by the structure of the propagators, see \equ(2.7) and
\equ(2.9), will have the form:
$$K_C=e^{ip_F(\sum_j \oo_j x_j)} \lis
K_{C}(\{x_j,\oo_j\}_{j=1,2,\ldots,|C|})\Eq(2.10)$$
where $x_j$ are vertices of the cluster $C$ to which the entering and
exiting lines are attached; the cluster may contain more vertices than
just the ones to which the external lines are attached: the positions
of such "extra" vertices must be considered as integration variables
(and as integrated), and a sum is understood to act over all the quasi
particles labels of the internal lines (consistent with the values of
the external lines $\oo_j$'s).

If $|C|=2,4$ we write the {\it Fourier transform} at $k=(k_0,\kk)$
of the kernels $\lis K_C(\ldots)$:

$$\eqalign{&
Z_1 2^{-1}\n^{(1)}_C
\d_{\oo_1,\oo_2}+Z_1(-i k_0 \z^{(1)}_C+\oo\kk \a^{(1)}_C)\,
\d_{\oo_1,\oo_2}+{\rm "remainder"}\cr
&Z_1^2\l^{(1)}_C\d_{\oo_1+\oo_2+\oo_3+\oo_4=0}+{\rm
"remainder"}\cr}\Eq(2.11)$$
where the first equation ($|C|=2$) is a function of one $k$ only while the
second equation ($|C|=4$) depends on four momenta $k$: one says that the
remainders are obtained by ``subtracting from the kernels their values
at the Fermi surface'' or by collecting terms tha {\it do not
conserve} the quasi particles momenta (like terms with
$\d_{\oo_2,-\oo_2}$ in the first equation or with $\oo_1+\ldots\oo_4\ne0$
in the second).

The remainder contains various terms which do not have the form of the
terms explicitly written in \equ(2.11): a form which could be as simple
as $\oo\cdot\kk\d_{\oo_1,-\oo_2}$ but that will in general be far more
involved.

In evaluating graphs we imagine, as described, them as made with clusters
and that the graph value is obtained by integrating the product of the
values of the kernels associated with the graph times the product of the
propagators of the lines that connect different clusters.

Furthermore we imagine to attach to each cluster with $2$ external
lines a label indicating that it contributes to the graph value
with the first term in the decomposition in \equ(2.11) only (which is the term
proportional to $\n^{(1)}$), or with the second term (which is proportional
to $(-i k_0\z^{(1)}+\oo\kk \a^{(1)}$) or with the remainder. This is easily
taken into account by attaching to the cluster an extra label $1,2$ or $r$.

Likewise we imagine to attach to each cluster with $4$ external lines a
label indicating that it contributes to the value with the first term in the
decomposition in \equ(2.11) only (which is the term proportional to
$\l^{(1)}$), or with the remainder.  This is again easily taken into account
by attaching to the cluster an extra label $1,r$.
\*

The label $r$ stands for ``remainder term'' or ``irrelevant term'',
however irrelevant does not mean neglegible, as usual in the
renormalization group nomenclature, (on the contrary they are in a way
the most important terms).

The next idea is to collect together all graphs with the same clusters
structure, \ie which become identical once the clusters with $2$ or
$4$ external lines are "shrunk" to points. Since the internal structure
of such graphs is different this means that we are collecting together
graphs of different perturbative order.

In this way we obtain a representation of the Schwinger functions that
is {\it no longer a power series} representation and the evaluation
rules for graphs in which {\it single vertex subgraphs} (or {\it single
node subgraphs}) with $2$ or $4$ external lines have a new
meaning. Namely a
four external lines vertex will mean a quantity $Z_1^2\l'$ equal to the
sum of $Z_1^2
\l_C^{(1)}$ of all the values of the clusters $C$ with $4$ external
lines and with label $1$.

The $2$ external lines nodes will mean

$$e^{i(\oo_1 \xx_1-\oo_2
\xx_2)p_F} \d_{\oo_1,\oo_2} Z_1 (\n' + \z'\dpr_t-i\a' \oo
\dpr_\xx)\d(x_1-x_2))\Eq(2.12)$$
where again $\n'$ or $\z',\a'$ are the sum of the contributions from
all the graphs with $2$ external lines and with label $0$ or $1$
respectively.
\*

It is convenient to define $\d'=\z'-\a'$ and to rewrite the $2$ external
lines nodes contributions to the product generating the value of a graph
simply as:

$$e^{i(\oo_1 \xx_1-\oo_2 \xx_2)p_F}\,
\d_{\oo_1,\oo_2}\, Z_1\,(2\,\n' + \d'\,\dpr_t+ \a'\,(\dpr_t-\oo\cdot\dpr_\xx)
)\,\d(x_1-x_2)\Eq(2.13)$$
One then notes that this can be represented graphically by saying that
$2$ external lines nodes in graphs which do not carry the label $r$ can
contribute in $3$ different ways to the product determining the graph
value. The $3$ ways can be distinguished by a label $0$, $1'$ and $z$
corresponding to the three addends in \equ(2.13).
\*

Any graph without $z$--type of nodes can be turned into a graph which
contains an arbitrary number of them, on each line connecting the
clusters.  And this amounts to saying that we can compute the series by
imposing that there is {\it not even a single vertex with two external
lines and with label $z$} simply by modifying the propagators of the
lines connecting the graphs: changing them from $Z_1^{-1} g^{(\le 0)}$
to $Z_0^{-1} g^{(\le 0)}$ with
$$Z_0=Z_1(1 +\a')\Eq(2.14)$$
This can be seen either elementarily by remarking that adding values
of graphs which contains chains of nodes with label $z$ amounts to
summing a geometric series (\ie precisely the series $\sum_{k=0}^\io
(-1)^k(\a')^k=(1+\a')^{-1}$ or, much more easily, by recalling that
the graphs are generated by a formal functional integral over
Grassmanian variables and checking \equ(2.14) from this remark without
any real calculation, see [BG95]. In the first approach care is needed
to get the correct relation
\equ(2.14) and it is wise to check it first in a few simple cases
(starting with the ``linear'' graphs which only contain nodes with one
entering line and one exiting line, see Fig. 1: the risk is to get
$Z_0=Z_1\,(1-\a')$ instead of \equ(2.14)).\annota{2}{\nota It is at
this point that using the sharply discontinuous $\ch$--functions would
cause a problem. In fact if one uses the smooth decomposition
\equ(2.14) is {\it no longer correct}: namely it would become $Z_0=
Z_1\,(1+\ch^{(0)}(\kk)\,\a')$ with the consequence that $Z_1$ would
no longer be a constant. At this point there are two possible ways out:
the first is to live with a $Z_0$ which dpends on $k$ and with the
fact that the quantities introduced below $Z_j$, $j\le-1$, will also
be $k$--dependent; this is possible but it is perhaps too different
from what one is used to in the phenomenological renormalization group
approaches in which quantities like $Z_j$ are usually constants. The
other possibility is to modify the propagator on scale $0$ from
$Z_0^{-1} g^{(\le0)}(\kk)$ to $g^{(\le0)}(\kk)\, (1+\a')/(1+
\ch^{(0)}(\kk)\,\a')$. The second choice implies that $g^{(\le h)}$
will no longer be exactly $\ch^{(h)}(\kk)/(-ik_0+\oo\cdot\kk)$ but it
will be gradually modified as $h$ decreases and the modification has
to be computed step by step. This is {\it also} unusual in the
phenomenological renormalization group approaches: the reason being
simply that in such approaches the decomposition with sharp
discontinuities is always used. The latter is not really convenient if
one wants to make estimates of large order graphs. Here this will not
be a poblem for us because we shall not do the technical work of
deriving estimates. In [BG90] as well as in [BGPS] the second choice
has been adopted.}

Correspondingly we set:

$$\n^{(0)}=2 \fra{Z_0}{Z_1}\n',\qquad
\d^{(0)}= \fra{Z_0}{Z_1}\d',\qquad
\l^{(0)}=  \fra{Z_0^2}{Z_1^2}\l'\Eq(2.15)$$

We can now {\it iterate the analysis}: we imagine writing the
propagators of the lines connecting the clusters so far considered and
that we shall call {\it clusters of scale $1$} as:

$$\fra1{Z_0} g^{(\le0)}=\fra1{Z_0} g^{(0)}+ \fra1{Z_0}
g^{(\le-1)}\Eq(2.16)$$
and proceed to decompose all the propagators of lines outside the
clusters of scale $1$ into propagators of scale $0$ or of scale $\le
-1$.

In this way, imagining all clusters of scale $0$ as points, we build a
new level of clusters (whose vertices are either vertices or clusters of
scale $0$): they consist of maximal sets of clusters of scale $1$ {\it
connected} via paths of lines of scale $0$.

Proceeding in the same way as in the above "step $1$" we represent the
Schwinger functions as sums of graph values of graphs built with
clusters of scale $0$ connected by lines with propagators on scale
$\le -1$ given by $Z_0^{-1}g^{(\le -1)}$ and with the clusters
carrying labels $1,2$ or $r$. Again we rearrange the $2$--external
lines clusters with labels $1,2$ as in \equ(2.13) introducing the
parameters $\d',\l',\a',\n'$ and graphs with nodes of type $z$ by
defining $\a'$ in an analogous way as the previous quantity with the
same name (relative to the scale $1$ analysis).

The one vertex nodes of such graphs with $2$ or $4$ external lines of
scale $\le -1$ will contribute to the product defining the graph value,
a factor $Z_{-1} 2\n^{(-1)}$ or $Z_{-1}\d^{(-1)}$ or $Z_{-1}^2
\l^{(-1)}$ (while the propagators in the clusters of scale $1$ and the
$2$ or $4$ nodes with two lines of scale $0$ emerging from them retain
the previous meaning). Again one sets:

$$Z_{-1}=Z_0(1 +\a')\Eq(2.17)$$
and correspondingly we set:

$$\n^{(-1)}=2 \fra{Z_{-1}}{Z_0}\n',\qquad
\d^{(-1)}=\fra{Z_{-1}}{Z_0}\d',\qquad
\l^{(-1)}=\fra{Z_{-1}^2}{Z_0^2}\l'\Eq(2.18)$$
and now we shall only have graphs with $2$ or $4$ external lines clusters
which carry a label $0,1'$ or $r$ as in the previous analysis of the
scale $1$ and the propagators connecting clusters of scale $0$ changed
from $Z_0^{-1} g^{(\le-1)}$ to $Z_{-1}^{-1} g^{(\le-1)}$.

Having completed the step $0$ we then ``proceed in the same way''
and perform ``step $-1$'' and so on.
\*
One can wander why the choice of the scaling factor $2$ in \equ(2.13)
and \equ(2.18) multiplying the ratio of the renormalization factors in
the definition of the new $\n_j$ or, for that matter, why the choice of
$1$ for the definition of the new $\d_j,\l_j$: these are dimensional
factors that come out naturally and any attempt at modifying the above
choices leads to a beta function, defined below, which is not uniformly
bounded as we remove the infrared cut--off. In other words: different
scalings can be considered but there is only one which is useful. It
could also be found by using arbitrary scaling and then look for which
one the estimates needed to get a convergent expansion can be made.
\*

The conclusion is a complete rearrangement of the perturbation expansion
which is now expressed in terms of graphs which bear various labels and,
most important, contain propagators that bear a scale index which gives
us information on the scale on which they are sizably different from
$0$. The procedure, apart from convergence problems, leads us to define
recursively a sequence $\l^{(j)},\d^{(j)},\n^{(j)},Z_j$ of constants
each of which is a sum of a formal power series involving values of
graphs with $2$ or $4$ external lines. The quantities $\V g_j=(\l^{(j)},
\d^{(j)},\n^{(j)})$ can be called the {\it running coupling constants}
while $Z_j$ can be called the {\it running wave function renormalization
constants}: here $j=1,0,-1,-2,\ldots$.

Of course all the above is nothing but algebra, made simple by the
graphical representation of the objects that we wish to compute. The
reason why it is of any interest is that, since the construction is
recursive, one derives expressions of the $\V g_j,Z_j$ in terms of the
$\V g_n,Z_n$ with $n>j$:

$$\eqalign{ &\fra{Z_{j+1}}{Z_j}= 1+ B'_j(\V g_{j+1},\V g_j,\ldots,\V
g_0)\cr &\V g_{j}=\L_j\,\V g_{j+1} +\V C_j(\V g_{j+1},\V g_j,\ldots,\V
g_0)\cr}\Eq(2.19)$$
where $\L_j$ is a matrix:

$$\L_j=\pmatrix{(Z_h/Z_{h-1})^2&0&0\cr
0&(Z_h/Z_{h-1})&0\cr
0&0&2 \, (Z_h/Z_{h-1})\cr}\Eq(2.20)$$
and the functions $B'_j, \V C_j$ are given by power series, so far
formal, in the running couplings. The expression of $Z_{j+1}/Z_j$ can
be used to eliminate such ratios in the second relation of \equ(2.19)
which therefore becomes

$$\eqalign{ &\fra{Z_{j+1}}{Z_j}= 1+ B'_j(\V g_{j+1},\V g_j,\ldots,\V
g_0)\cr &\V g_{j}=\L\,\V g_{j+1} + \V B_j(\V g_{j+1},\V g_j,\ldots,\V
g_0)\cr}\Eq(2.21)$$
where $\L$ is the diagonal matrix with diagonal $(1,1,2)$. The scalar
functions $B'_j$ and the three components vector functions $\V
B_j=(B_{j,1},B_{j,2},B_{j,3})$ are called the {\it beta functional} of
the problem.

There are two key points, which are nontrivial at least if compared to
the above simple algebra and which we state as propositions
\*

\0{\bf Proposition 1 \it (regularity and boundedness of the beta function):
Suppose that there is $\e>0$ such that $|\V g_j|<\e,
|Z_j/Z_{j-1}\,-\,1|<\e$ for all $j\le 1$ then if $\e$ is small enough
the power series defining the beta functionals converge. Furthermore
the functions $B_j,B'_j$ are uniformly bounded and have a dependence
on the arguments with label $j+n$ exponentially decaying as $n$ grows,
namely there exist constants $D,\k$ such that if $\V G=(\V
g_{j+1},\ldots, \V g_0)$ and $\V G'= (\V g'_{j+1},\ldots, \V g'_0)$
with $\V G$ and $\V F$ differing only by the $(j+n)$--th ``component''
$\V d=\V g'_{j+n}-\V g_{j+n}\ne \V0$, then for all $j\le0$ and all
$n\ge0$

$$\eqalign{
&|B_j(\V G)|,\ |B_j'(\V G)|\le D \e^2\cr
&|B_j(\V G')-\V B_j(\V G)|,\
|B'_j(\V G')-\V B'_j(\V G)|\le D\, e^{-\k n}\, \e\,|\V d|}\Eq(2.22)$$
if $\e$ is small enough: \ie the ``memory'' of the ``beta functionals''
$B_j,B'_j$ is short ranged.
\\
The Schwinger functions are expressed as convergent power series in
$\V g_j$ in the same domain $|\V g_j|<\e$.}
\*

The difficult part of the proof of the above proposition is to get the
convergence of the series under the hypotheses $|\V g_j|<\e$,
$|Z_j/Z_{j-1}-1|<\e$ for all $j$: this is possible because the system is
a fermionic system and one can collect the contributions of all
graphs of a given order $k$ into a few, \ie not more than an
exponential in $k$, groups each of which gives a contribution that {\it is
expressed as a determinant which can be estimated without really
expanding it} into products of matrix elements (which would lead to
bounding the order $k$ by a quantity growing with $k!$) by making use of
the Gram--Hadamard inequality. Thus the $k!^{-1}$ that is in the
definition of the values compensates the number of labels that one can
put on the trees and the number of Feynman graphs that is also of order
$k!$ is controlled by their representability as determinants that can be
bounded without generating a $k!$ via the Hadamard inequality. The basic
technique for achieving these bounds is well established after the work
[Le87]. A second non trivial result is
\*

\0{\bf Proposition 2 \it (short range and  asymptotics of the beta
function):
\\
Let $\V G^0=(\V g,\V g,\ldots,\V g)$ with $\V g=(\l,\d,\n)$ then the
function $\V B_j(\V G^0)$ defines an analytic function of $\V g$, that
we shall call ``beta functional'', by setting

$$\V\b(\V g)=\lim_{j\to-\io} \V B_j(\V G^0)\Eq(2.23)$$
for $|\V g|<\e$.  The limit is reached exponentially $|\V\b(\V g)- \V
B_j(\V G^0)|< \e^2 D e^{-\k |j|}$, for some $\k>0, D>0$ provided $|\V
g|<\e$.}
\*

Finally the key result, [BG90], [BGPS], is
\*

\0{\bf Proposition 3 \it(vanishing of the beta function): If $\V g=(\l,\d,0)$
then the functions $\V \b(\V g)=\V 0$ provided $|\V
g|<\e$. Furthermore for some $D,\k>0$ it is, for all $j\le0$

$$\eqalign{
&B_{j3}(\V g_{j+1},\ldots,\V g_0)=\n_{j+1}\l_{j+1}^2 B'_{j3}(\V
g_{j+1},\ldots,\V g_0)+ e^{\k j} B''_{j3}(\V
g_{j+1},\ldots,\V g_0)\cr
&|B'_{j3}(\V g_{j+1},\ldots,\V g_0)|< D,\qquad
|B''_{j3}(\V g_{j+1},\ldots,\V g_0)|< D\,\e^2\cr}\Eq(2.24)$$
provided, for $h=0,\ldots,j+1$, $|\V g_h|<\e$.}
\*

The above propositions are proved in [BG90], [BGPS], [BM00].  The
vanishing of $\V\b(\l,\d,0)$ is proved in a rather indirect way. We
proved that the function $\V \b$ is the {\it same} for the model
\equ(2.1) and for a similar model, the {\it Luttinger model}, which is
exactly soluble; but which can be also studied with the technique
described above: and the only way the exactly soluble model results
could hold is to have $\V\b=\V0$.

The vanishing of the beta function seems to be a kind of Ward
identity: it is easy to prove it directly if one is willing to accept
a formal proof. This was pointed out, after the work [BG90], in other
papers and it was believed to be true probably much earlier in some
equivalent form, see [So79]; note that the notion of the beta function
is {\it intrinsic to the formalism} of the renormalization group and
therefore a precise conjecture on it could not even be stated before
the '70s; but of course the existence and importance of infinitely
many identities had already been noted.

Given the above propositions one shows that ``things go as if'' the
recursion relation for the running couplings was, up to exponentially
small corrections, a simple memoryless evolution $\V g_{j-1}=\V\b(\V
g_j)+O(e^{-\k|j|})$: the propositions say in a precise way that this
is asymptotically, as $j\to-\io$, true. This tells us that the running
couplings $\l_j,\d_j$ stay constant (because $\b_1,\b_2$ vanish):
however they in fact tend to a limit as $j\to-\io$ exponentially fast
because of the corrections in the above propositions, provided we can
guarantee that also $\n_j\tende{j\to-\io}0$ exponentially fast and
that the limits of $\l_,,\d_j$ do not exceed $\e$ (so that the beta
functionals and the beta function still make sense).

It is now important to recall that we can adjust the initial value of
the chemical potential.\annota{3}{\nota Which is a ``{\it relevant
operator}'', in the sense that if regarded as a running coupling it is
roughly multiplied by $2$ at each change of scale, \ie $\n_{j-1}\sim 2
\n_j$.} This freedom corresponds to the possibility of changing the
chemical potential `correction'' $\n$ in \equ(2.1) and tuning its
value so that $\n_h\to 0$ as $h\to-\io$.

Informally if $\n_0$ is chosen ``too positive'' then $\n_j$ will grow
(exponentially) in the positive direction (becoming larger than $\e$,
a value beyond which the series that we are using become meaningless);
if $\n_0$ is chosen ``too negative'' the $\n_j$ also will grow
(exponentially) in the negative direction: so there is a unique choice
such that $\n_j$ can stay small (and, {\it actually}, it can be shown
to converge to $0$.\annota{4}{\nota A simplified analysis is obtained
by ``neglecting memory corrections'' \ie using as a recursion relation
$\V g_j=\L \V g_{j+1}+\V\b(\V g_{j+1})$ with $\V\b(\V g)$ verifying
\equ(2.24): this gives that $\l_j,\d_j\tende{j\to-\io}
(\l_{-\io},\d_{-\io})$ exponentially fast
and $\n_j\tende{j\to-\io}0$ exponentially fast provided $\n_0$ is
suitably chosen in terms of $\l_0,\d_0$: otherwise everything
diverges.}

The vanishing of the beta function gives us the existence of a
sequence of running couplings $\V g_j=(\l_j,\d_j,\n_j)$ which converge
exponentially fast to $(\l_{-\io},\d_{-\io},0)$ as $j\to-\io$ if
$\n_0$ are conveniently chosen: and one can prove that
$\l_{-\io},\d_{-\io},\n_0$ are analytic in $\l$ for $\l$ small enough,
[BGPS].

In this way one gets a convergent expansion of the Schwinger functions:
which leads to an essentially complete theory of the one dimensional
Fermi gas with spin zero and short range interaction.
\*

\0{\bf\S3. The conceptual scheme of the renormalization group approach
followed above.}
\numsec=3\numfor=1\*

The above schematic exposition of the method is a typical example of
how one tries to apply the multiscale analysis that is commonly called
a ``renormalization group approach'':

(1) one has series that are easily shown to be finite order by order
possibly provided that some free parameters are suitably chosen
(``formal renormalizability theory'': this is the proof in [LW60] that if
$\n_h$ in \equ(2.2) are suitably chosen we obtain a well defined
perturbation series in powers of $\l$.

(2) However the series even when finite term by term come with poor
bounds which grow at order $k$ as $k!$ which, nevertheless are often non
trivial to obtain (although this not so in the case \equ(2.1) discussed
here unlike the case discussed in the next sections).

(3) One then tries to reorganize the series by leaving the original
parameters ($\l,\n$) in the present case as $\m$ is fixed) as
independent parameters and collecting terms together. The aim being to
show that they become very {\it convergent power series in a sequence
of new parameters}, the ``running couplings'' $\n^{(h)},\d^{(h)}$ and
$\l^{(h)}$ in the present case, {\it under the assumption that such
parameters} are small (they are functions, possibly singular, of the
initial parameters of the theory, $\l,\n$ in the case \equ(2.1), as
$\d$ has to be imagined to be $0$).

(4) The running couplings, essentially by construction, also verify a
recursion relation that makes sense {\it again} under the assumption
that the parameters are small. This relation allows us to express (if it
makes sense) successively the running couplings in terms of the
preceding ones: the running couplings are ordered into a sequence by
``scale labels'' $h=1,0,-2,\ldots$. The recursion relation is
interpreted as an evolution equation for a dynamical system (a map
defined by the beta function(al)): it generates a ``renormalization
group trajectory'' (the sequence $(\l_h,\d_h,\n_h)$ out of the original
parameters $\l,\n$ present in \equ(2.1), as $\d$ has to be taken as
$0$).

(5) One then shows that {\it if the free parameters in the problem},
(\ie $\l,\n$ in \equ(2.1)) {\it are conveniently chosen}, then the
recursion relation implies that the trajectory stays bounded and
small, thus giving a precise meaning to \equ(2.2)), and actually the
limit relation holds
$(\l^{(h)},\d^{(h)},\n^{(h)})\tende{h\to-\io}(\l_\io,\d_\io,0)$ (this
is achieved in the above Fermionic problem by fixing $\n$ as a
suitable function of $\l$, see [BGPS].

(6) Hence the whole scheme is self--consistent and it remains to check
that the expressions that one thus attributes to the sum of the series
are indeed solutions of the problem that has generated them: not
unexpectedly this is the easy part of the work, because we have always
worked with formal solutions which ``only missed, perhaps, to be
convergent''.

(7) The first step, \ie going to scale $0$ is different from the
others as the propagators have no ultraviolet cut off (see the graph
of $\ch^{(1)}$ in Fig. 2). Although there are no ultraviolet
divergences the control of this first step offers surprising
difficulties (due to the fact that in the direction of $k_0$ the decay
of the propagators is slow making various integrals improperly
convergent): the analysis is done in [BGPS] and [GS93].

Note that the above scheme leaves room for the possibility that the
running couplings rather than being analytic functions of a few of the
initial free parameters are singular: this does not happen in the above
fermionic problem because some components of the beta function vanish
identically: this is however a peculiarity of the fermionic models. In
other applications to field theory, and particularly in the very first
example of the method which is the hierarchical model of Wilson, this is
by far not the case and the perturbation series are {\it not analytic}
in the running couplings but {\it just asymptotic} in the actual free
parameters of the theory. The method however ``reduces'' the
perturbation analysis to a recursion relation in small dimension (namely
$3$ in the case \equ(2.1)) which is also usually easy to treat
heuristically. The $d=2$ ground state fermionic problem (\ie \equ(2.1)
in $2$ space dimensions) provides, however, an example in which even the
heuristic analysis is not easy.
\*

In the following section we discuss another problem where the beta
function does not vanish, but one can guarantee the existence of a
bounded and small solution for the running couplings thanks to a ``gauge
symmetry'' of the problem. This is an interesting case as the theory has
{\it no free parameters} so that it would not be possible to play on
them to find a bounded trajectory for the renormalization group running
constants. This also illustrates another very important mechanism that
can save the method in case there seemed to be no hope for its use,
namely a symmetry that magically eliminates all terms that one would
fear to produce ``divergences'' in formal expansions. Again the case
studied is far from the complexity of gauge field theory because it
again leads to the result that the perturbation series itself is
summable (unlike gauge field theories which can only yield asymptotic
convergence): but it has the advantage of being a recognized difficult
problem and therefore is a nice illustration of the role of symmetries
in the resummation of (possibly) divergent series and the power of the
renormalization group approach in dealing with complex problems.
\*

\0{\bf\S4. The KAM problem.}
\numsec=4\numfor=1\*

Consider $d$ rotators with angular momentum $\V A=(A_1,\ldots, A_d)\in
R^d$ and positions $\V\a =(\a_1,\ldots,\a_d)\in T^d=[0,2\p]^d$; let
$J>0$ be their inertia moment and suppose that $\e f(\V\a)$ is the
potential energy in the configuration $\V \a$, which we suppose to be an
even trigonometric polynomial (for simplicity) of degree $N$.  Then the
system is Hamiltonian with Hamiltonian function

$$\HH=\fra1{2J} \V A^2+\e f(\V \a)\Eq(4.1)$$
giving rise to a model called ``Thirring model''.\annota{5}{\nota (1)
The global canonical transformations $\CC$ of $R^d\times T^d$ with
generating functions
$S(\AA,\aa)=N\AA'\cdot\aa+\V\g(\aa)\cdot\AA'+\f(\aa)$ parameterized by
an integer components non singular matrix $N$, and analytic functions
$\V g(\aa),f(\aa)$ leave invariant the class of Hamiltonians of the
form $H=(\AA, M(\aa)\AA)/2 + \AA\cdot\V g(\aa)+f(\aa)$. The subgroup
$CL_d(R)$ of the global canonical coordinate trransformations $\CC$
was (remarked and) used by Thirring so that \equ(4.1) is called the
``Thirring model'', see [Th83].
\\
(2) The function $\V H_\e(\pps)$ in \equ(4.2) must have zero average
over $\pps$ or, if $\pps\to\pps+\oo_0 t$, over time: hence the surviving
quasi periodic motions can be parameterized by their spectrum $\oo_0$
or, equivalently, by their average action $\AA_0= J\oo_0$.. The
``spectral dispersion relation'' between the average action $\AA_0$ and
the frequency spectrum is {\it not twisted} by the
perturbation. Furthermore the function $\e_(\oo_0,J)$ can be taken
monotonically increasing $J$: $J^{-1}$ is called the ``{\it twist
rate}. The latter two properties motivated the name of ``{\it twistless
motions}'' given to the quasi periodic motions of the form
\equ(4.2) for Hamiltonians like \equ(4.1).
\\
(3) The invariance under the group $CL_d(R)$ has been used widely in the
numerical studies of the best treshold value $\e(\oo_0,J)$ and a deeper
analysis of this group would be desirable, particularly a theory of its
unitary representations.}

For $\e=0$ motions are quasi periodic (being $t\to (\V A_0,\V\a_0+\oo_0
t)$ with $\oo_0= J^{-1}\V A_0$) and their ``spectrum'' $\oo_0$ fills the
set $S_0\= R^d$ of {\it all} vectors $\oo_0$: there is a $1$-to-$1$
correspondence between the spectra $\oo_0$ and the angular momenta $\V
A_0$.
\*

{\it Question: If $\e\ne0$ can we find, given $\oo_0\in S_0$ a
perturbed motion, \ie a solution of the Hamilton equations of
\equ(4.1), which has spectrum $\oo_0$ and, as $\e\to0$, reduces with
continuity to the unperturbed motion with the same spectrum? or less
formally: which among the possible spectra $\oo\in S_0$ survives
perturbation?}

\*
Analytically this means asking whether two functions $\V H_\e(\pps),\V
h_\e(\pps)$ on $T^d$ exist, are divisible by $\e$ and are such that if
$\V A_0= J\V \o_0$ and if we set

$$\eqalign{
\AA=&\AA_0+\V H_\e(\pps)\cr
\aa=&\pps+\V h_\e(\pps)\cr},\qquad \pps\in T^d\Eq(4.2)$$
then $\pps\to\pps+\oo_0 t$ yields a solution of the equations of motion
for $\e$ small enough.

It is well known that {\it in general} only ``non resonant'' spectra can
survive: for instance ({\it KAM theorem}) those which verify, for some
$\t,C>0$

$$|\oo_0\cdot\nn|^{-1} < C |\nn|^\t\qquad \forall \nn\in
Z^d=\{integer\
vectors\}, \ \nn\ne \V0\Eq(4.3)$$
for some $C,\t>0$ ({\it Diophantine vectors}) and we restrict for
simplicity to such vectors. Furthermore given $\oo_0$ the perturbation
size $\e$ has to be small enough: $|\e|\le
\e_0(\oo_0,J)$.

To find $\V H_\e,\V h_\e$ we should solve the equation (setting $J\=1$),

$$(\oo_0\cdot\Dpr_\pps)^2 \,\V h(\pps)=-\e\,(\Dpr_\aa f)(\pps+\V
h(\pps))\Eq(4.4)$$
and if such $\V h$ is given then, setting $\V h_\e(\pps)=\V h(\pps),\V
H_\e(\pps)=(\oo_0\cdot\Dpr)\,\V h(\pps)$ one checks that
\equ(4.2) has the wanted property (\ie $\pps\to\pps+\oo_0 t$ is a
motion for \equ(4.1)).
\*

To an exercized eye \equ(4.4) defines the expectation value of the
$1$--particle Schwinger function of the euclidean field theory on the
torus $T^d$ for two vector fields $\V F^\pm(\pps)=(F^\pm_1(\pps),\ldots,
F^\pm_{d}(\pps))$ whose free propagator is
$$\media{F^\s_\ell(\pps)\, F^{\s'}_m(\pps')}=
\d_{\s,-\s'}\,\d_{\ell,m}
\fra{-1}{(2\p)^d}\sum_{\nn\ne\V
0}\fra{e^{i(\pps-\pps')\cdot\nn}}{(\oo_0\cdot\nn)^2}
\Eq(4.5)$$
and the interaction Lagrangian is, see [Ga95]

$$\LL(F)=\e\ig_{T^d}\V F^+(\pps)\cdot\Dpr_\aa(\pps+\V
F^-(\pps))\, d\pps\Eq(4.6)$$
If $P_0(d\V F)$ the ``functional integral'' defined by Wick's
rule with propagator \equ(4.5) then

$$\V h(\pps)=\fra{\ig \V F^-(\pps)\, e^{\e\ig_{T^d}\V
F^+(\pps)\cdot\Dpr_\aa f(\pps+\V F^-(\pps))}\,P_0(d\V F)}
{\ig e^{\e\ig_{T^d}\V
F^+(\pps)\cdot\Dpr_\aa f(\pps+\V F^-(\pps))}\,P_0(d\V F)}\Eq(4.7)$$

At first sight this is a ``sick field theory''. Not only the fields $\V
F^\pm(\pps)$ do not come from a positive definite propagator, hence
\equ(4.7) has to be understood as generating a formal expansion in $\e$
of $\V h$ with integrals over $\V F$ being {\it defined} by the Wick
rule, but also the theory is {\it non polynomial} and naively non
renormalizable.

The ``only'' simplification is that the Feynman diagrams of \equ(4.7)
are (clearly) tree--graphs, \ie loopless: this greatly simplifies the
theory which, however, remains non renormalizable and non trivial
(being equivalent to a non trivial problem).

It is not difficult to work out the Feynman rules for the diagrams
expressing the $k$--th order coefficient of the power series expansion
in $\e$ of \equ(4.7).

Consider a rooted tree with $k$ nodes: the branches are considered
oriented towards the root which is supposed to be reached by a single
branch and which is not regarded as a node of the tree (hence the
number of nodes and the number of branches are equal).

\*

\eqfig{200pt}{141.6pt}{
\ins{-29pt}{75pt}{\it root}
\ins{0pt}{71pt}{$\nn\kern-3pt=\kern-3pt\nn(v_0)$}
\ins{0.pt}{91pt}{$\V e$}
\ins{50pt}{71pt}{$v_0$}
\ins{46pt}{91pt}{$\nn_{v_0}$}
\ins{127pt}{100.pt}{$v_1$}
\ins{121.pt}{125.pt}{$\nn_{v_1}$}
\ins{92.pt}{41.6pt}{$v_2$}
\ins{158.pt}{83.pt}{$v_3$}
\ins{191.7pt}{133.3pt}{$v_5$}
\ins{191.7pt}{100.pt}{$v_6$}
\ins{191.7pt}{71.pt}{$v_7$}
\ins{191.7pt}{-8.3pt}{$v_{11}$}
\ins{191.7pt}{16.6pt}{$v_{10}$}
\ins{166.6pt}{42.pt}{$v_4$}
\ins{191.7pt}{54.2pt}{$v_8$}
\ins{191.7pt}{37.5pt}{$v_9$}
}{bggmfig0}{}
\kern0.9cm
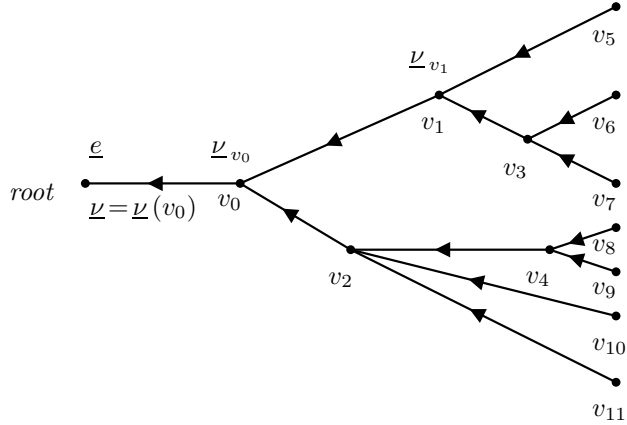
\didascalia{Fig. 5: A graph $\th$ with
$p_{v_0}=2,p_{v_1}=2,p_{v_2}=3,p_{v_3}=2,p_{v_4}=2$ and $k=12$, and some
labels. The line numbers, distinguishing the lines are not shown. The
lines length should be the same but it is drawn of arbitrary size.  The
momentum flowing on the root line is $\nn=\nn(v_0)=$ sum of all the
nodes momenta, including $\nn_{v_0}$.}

We attach to each node (or ``vertex'') $v$ of the tree a vector
$\nn_v\in Z^d$, called a ``mode label'', and to a line oriented from
$v$ to $v'$ we attach a ``current'' or ``momentum'' $\nn(v)$ and a
``{\it propagator}'' $g(\nn(v))$:

$$\nn(v)\defi \sum_{w\le v} \nn_v, \qquad g(\oo_0\cdot\nn)\defi
(\oo_0\cdot\nn)^{-2}\Eq(4.8)$$
if $v$ is the ``first node'' of the tree, see $v_0$ in Fig. 5,
then the momentum $\nn(v)$ is called the total momentum of the tree.

One defines the {\it value} ${\rm Val}\,(\th)$  of a tree $\th$
decorated with the described labels as

$${\rm Val}(\th)=k!^{-1}\,\prod_{v\in {\rm nodes}}
f_{\nn_v}\fra{\nn_v\cdot\nn_{v'}}{(\oo_0\cdot\nn(v))^2}\Eq(4.9)$$
where $f_\nn$ are the Fourier coefficients of the perturbation
$f(\aa)=\sum_\nn f_\nn e^{i\nn\cdot\pps}$, with $|\nn|\le N$,
$f_{\nn}=f_{-\nn}$, and $v'$ denotes the node following $v$ and, if
$v$ is the first node of the tree (so that $v'$ would be the root
which (by our conventions) is not a vertex) then $\nn_{v'} $ is some
unit vector $\V e$, see Fig. 3.

Given the above Feynman rules, which one immediately derives from
\equ(4.6), \equ(4.7) the component along the vector $\V e$, labeling
the root, of the $k$--the order Fourier coefficient $\V h^{(k)}_\nn$
of the function $\V h(\pps)\cdot\V e$, which we write as

$$\V h_\nn\cdot\V e=\e\V h_\nn^{(1)}\cdot\V e+\e^2\V
h_\nn^{(2)}\cdot\V e
+\ldots,\Eq(4.10)$$
{\it is simply the sum of the values ${\rm Val}\th$ over all trees} $\th$
which have total momentum $\nn$, $k$ nodes, and {\it no} branch
carrying $\V0$--momentum.

One can check directly that $\V h(\pps)$ so defined is a formal solution
of \equ(4.4): the series \equ(4.10) with the coefficients defined as
above is called the {\it Lindstedt series} of the KAM problem: it was
introduced, at least as a method for computing the low order
coefficients $\V h^{(k)}$, by Lindstedt and Newcomb in celestial
mechanics problems and it was shown to be possible to all orders by
Poincar\'e (one has to show that the algorithm generating the series
does not produce graphs with branches carrying $\V0$ momentum which, by
\equ(4.9), would yield meaningless expressions for the corresponding
tree values), see [Po92], vol. 3.

The number of trees of order $k$ that do not differ only by the labeling
of the lines (\ie that are topologically the same) is bounded
exponentially in $k$ while the total number of trees (labels included)
is, therefore, of order $k!$ times an exponential in $k$.

Therefore taking into account the $k!^{-1}$ in \equ(4.8) we see that
the perturbation series might have convergence problems only if there
exist individual graphs whose value is too large, \eg $O(k!^\g)$ for
some $\g>0$.

Such graphs {\it do exist}; an example:

\eqfig{230pt}{70pt}{
\ins{25pt}{20pt}{$\st1$}
\ins{45pt}{-6pt}{$\st-\nn_0$}
\ins{25pt}{40pt}{$\st\nn_0$}
\ins{55pt}{20pt}{$\st3$}
\ins{75pt}{-6pt}{$\st-\nn_0$}
\ins{55pt}{40pt}{$\st\nn_0$}
\ins{85pt}{20pt}{$\st5$}
\ins{105pt}{-6pt}{$\st-\nn_0$}
\ins{85pt}{40pt}{$\st\nn_0$}
%\ins{120pt}{20pt}{$4$}
\ins{170pt}{20pt}{$\st \fra{2k}3+1$}
\ins{170pt}{45pt}{$\st \nn_{1}$}
\ins{160pt}{-6pt}{$\st-\nn_0$}
\ins{140pt}{40pt}{$\st\nn_0$}
\ins{220pt}{0pt}{$\st k/3$}
\ins{205pt}{-6pt}{$\st \nn_{k/3}$}
\ins{205pt}{65pt}{$\st \nn_2$}
}
{xyz}{}
\*\*
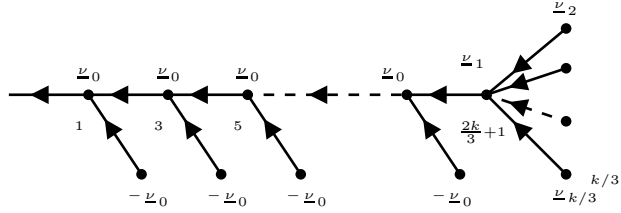
\0{\it Fig. 4: \nota A tree of order $k$ with momentum
$\nn=\nn_1+\ldots\nn_{k/3}$ and value of size of order $(k/3)!^\t$ if
$\nn=\nn_1+\ldots\nn_k$ is ``as small as it can possibly be'': or is
``almost resonant'' \ie such that $\oo_0\cdot\nn\sim
C^{-1}|\nn|^{-\t}$. The tree has $k/3\,+\,1$ branches carrying
momentum $\nn=\sum_{i=1}^{k/3}\nn_i$, \ie the $k/3\,+\,$ horizontal
branches. The last $k/3\,-1$ branches have momenta $\nn_i,
i=2,\ldots,k/3$ so arranged that their sum plus $\nn_1$, \ie $\nn$, is
``almost resonant''.}
\*

Therefore even though the theory is loopless and its
perturbation series is well defined to all orders, yet it is non
trivial because the $k$--th order might be ``too large''.

This is a typical situation in ``infrared divergences'' due to too
large propagators: in fact the ``bad'' graphs (like the one in Fig. 5)
are such because $(\oo\cdot\nn)^{-2}$, \ie the propagator, is too
large.
\*

It is remarkable that the same strategy used in the analysis of the
Fermi gas theory, \ie the renormalization group approach outlined in
general terms in \S3, works in this case. We decompose the
propagator as

$$\fra1{(\oo_0\cdot\nn)^{2}}=\sum_{h=-\io}^1
\fra{\ch^{(h)}(\oo_0\cdot\nn)}{(\oo_0\cdot\nn)^{2}}=
\sum_{h=-\io}^1 2^{-2h} g^{(h)}(2^h \oo_0\cdot\nn)\Eq(4.11)$$
where if $h\le 0$ we have set $\ch^{(h)}(x)\defi \ch(2^{-h}x)$ with
$\ch(x)=0$ unless $x$ is in the interval $2^{-1}< |x|\le 1$ where
$\ch(x)\=1$, and $\ch^{(1)}$ is defined to be identically $1$ for
$|x|\ge1$ and $0$ otherwise so that $1\=\sum_{h=-\io}^1 \ch^{(h)}(x)$:
see Fig. 2 and \equ(2.9).\annota{6}{\nota In this problem there is
{\it no need} to use functions $\ch^{(h)}$ which are not as in Fig. 4,
\ie with smoothed out discontinuities: there is, however, a minor
difficulty also in this case. The decomposition above,
\equ(4.11), can be done exactly as written only if $\oo_0\in R^d$ is
outside a certain set of zero volume in $R^d$. Although already the
Diophanntine property \equ(4.3) holds only outside a set of zero
volume in $R^d$, as is well known, the set of $\oo_0\in R^d$ for which
what follows can be done literally as described is slightly smaller
(although {\it still} with a complement of zero volume). However the
following discussion can be repeated under the only condition that
$\oo_0$ verifies \equ(4.3) provided one does not insist in taking a
sequence of scales that are exactly equal to $2^h$, $h=1,0,-1,\ldots$
and one takes a sequence of scales that have bounded but suitably
variable successive ratios. Here we ignore this problem, see [Ga94]
for the cases that work as described here (first reference) and for
the general case (second reference) discussed under the only
``natural'' condition \equ(4.3).}
\*

Given a Feynman graph, \ie a tree $\th$ (with the decorating labels)
one replaces $g(\oo_0\cdot\nn)$ by the last sum in \equ(4.11) and we
obtain trees with branches bearing a ``{\it scale label}''.

We collect the graph lines into clusters of scale $\ge h$ with at
least one line of scale $h$ and define the order $n$ of the
cluster to be the number of nodes in it.

It is not difficult, see [P\"o84] and [Ga94], to see that {\it only graphs
which contain one incoming and one outgoing line with the same momentum}
(hence same scale) are the source of the problem: for instance in the
case of the graph in Fig. 4 all horizontal lines have the same momentum
of scale $h\simeq \log k^{-\t}$ while all the other lines have scale
$O(1)$ because $|\nn_j|\le n$. The subgraphs that contain one incoming
and one outgoing line with the same momentum of scale $h\le 0$ have been
called in [El96] {\it resonances}, perhaps not very appropriately given
the meaning that is usually associated with the word resonance but we
adopt here the nomenclature of the latter breakthrough work. The
subgraphs which would be resonances but have maximal scale $h=1$ are not
considered resonances.

Therefore it is natural to collect together all graphs which contain
{\it chains} of $1,2,3\ldots$ clusters with equal incoming and
outgoing momenta on scale $h\le0$. For instance the graph in Fig. 4
contains a chain of $k/3$ clusters, each containing a single line
(namely the lines with $\nn_0,-\nn_0$ modes at their extremes), and
one cluster with $k/3-1$ lines (the lines entering the node $2k/3+1$)
and $k/3+1$ lines external to the resonant clusters, the horizontal
lines.  Call

$$\eqalign{
\G^{(h)}(\nn) =&\, \hbox{sum of $\e^{n}$ times the value of the
clusters with $n$ nodes and single}
\cr&\hbox{ incoming and outgoing lines of
equal momentum and scale $h$}\cr}\Eq(4.12)$$
which can be given meaning because disregarding the incoming line we
can regard the subcluster with one entering and one exiting line as a
tree $\th$ (or subtree) with root at the node where the exiting line
ends, so that {\it its value $\G^{(h)}(\nn)$ will be naturally defined}
as the value of $\th$ times $(\oo_0\cdot\nn)^{-2}$ (which takes into
account the propagator of the line entering the cluster).

\*
This leads to a rearrangement of the series for $\V h$ in which the
propagator of a line with momentum $\nn$ on scale $h$ is $\G^{(h)}(\nn)$
rather than $(\oo_0\cdot\nn)^{-2}\ch^{(h)}(\oo_0\cdot\nn)$ and there are
{\it no more} clusters with one incoming and one outgoing line of equal
momentum.

Of course the graphs with $k$ nodes give a contribution to $\V h$
which is no longer proportional to $\e^k$ because they contain
quantities $\G^{(h)}(\nn)$ which are (so far formally) power series in
$\e$.

The same argument invoked above, [P\"o], [Ga94], gives again that
{\it if for some constant $R$ and for all $\e$
small enough one could suppose that}

$$|\G^{(h)}(\nn)|\le R\, (\oo_0\cdot\nn)^{-2}\Eq(4.13)$$
{\it then the series for $\V h_\nn$ would be {\it convergent} for $\e$ small
enough}.

Furthermore if \equ(4.13) holds for scales $0,-1,\ldots, h+1$ then one
sees that there exist $a_h,b_h,r_h$ such that

$$\G^{(h)}(\nn)=
\fra{a_h}{(\oo_0\cdot\nn)^{4}}+\fra{b_h}{(\oo_0\cdot\nn)^3}+
(1+r_h(\nn)) \lis g^{(h)}(\oo_0\cdot\nn)\Eq(4.14)$$
with $|r_h|<R\e,\,|\lis g^{(h)}(\oo_0\cdot\nn)|< N^2
(\oo_0\cdot\nn)^{-2}$, and we see therefore that $a_h,b_h$ play the role
of ``{\it running couplings}'': they are non trivial functions of $\e$
but {\it they are the only quantities to control} because if we can show
that they are bounded by $\e R$, say, then the convergence of the
perturbation series would be under control because {\it we are reduced
essentially to the case in which no graphs with resonances are present}.
\*

Naturally, by the principle of conservation of difficulties, the
quatities $a_h$ and $b_h$ are given by power series in $\e$ whose
convergence is \ap difficult to ascertain.

As in the case of the Fermi gas, setting $\V c_h=(a_h,b_h)$ the very
definition of such constants in terms of sums of infinitely many
Feynman graphs implies that they verify a recursion relation

$$\V c_h=\L \V c_{h+1}+ \V \BB_h(\V c_{h+1},\ldots,\V c_{0}), \qquad
h\le0\Eq(4.15)$$
where $\L$ is a $2\times2$ diagonal matrix with diagional elements
$2^2,2$, and $\V c_1\=\V0$ because by definition $\G^{(h)}\ne g^{(h)}$
only for $h\le0$. The analogy with the previous Fermi system problem
seems quite strong. The function $\V \BB_h(\{\V c\})_{h+1,\ldots,0}$
{\it is not homogeneous in the $\V c_{h}$}: and this is an important
difference with respet to the previous fermionic case.

Clearly if $a_h,b_h\ne0$ for some $h$ we are ``lost'' because
although, {\it under the boundedness condition} \equ(4.13), on the
running couplings the beta function $\BB_h$ is well defined by a
convergent series, and although the whole rearranged perturbation
series is convergent under the same condition, we shall not be able to
prove that the running couplings stay small so that \equ(4.13) is
selfconsistent. In fact {\it both} $a_h,b_h$ are ``relevant
couplings'' (because the elements of the matrix $\L$ are $>1$) and the
two data $a_1,b_1$ (which are $0$ for $h=1$ because, by definition,
there are no resonances of scale $1$) will need to be carefully tuned
so that the renormalization group trajectory $a_h,b_h$ that \equ(4.15)
generates from $a_1,b_1$ is bounded: {\it there are however no free
parameters to adjust in \equ(4.1)!}

The situation is very similar to the one met in the Fermi liquid
theory: in that case one solves the problem by showing that the beta
function vanishes, at least asymptotically, for the marginal
couplings $\l_h,\d_h$: it remains the relevant coupling $\n_h$ which
can be bounded only because we have in that problem the freedom of
``adjusting'' a free parameter (the chemical potential $\n$).

In the present case we have two relavant parameters, $a_h,b_h$, and {\it
no free parameter} in the Hamiltonian: the situation would be hopless
{\it unless} it just happened that the correct initial data for a
bounded renormalization group trajectory were precisely the ones that we
have, namely $a_1=b_1=0$. This means that one should prove the identity
$\BB_h(\V 0)\=\V0$ for all $h\le0$ so that $\V c_h\=\V0$ for all
$h\le1$.
\*

The {\it vanishing} of the beta function at $\V c=\V 0$ was understood
in [El96] and it can be seen in various ways, see also [Ga94]. It is
however always based on symmetry properties of the model (the choice of
the origin of $\pps$ plays the role of a ``gauge symmetry'': the
interpretation of the cancellations used by [El96] as a consequence of a
gauge symmetry was pointed out and clearly stated in [BGK99], although
of course the symmetry is used in the analysis in [El96] and in great
detail in [Ga94]).

We are back to a very familiar phenomenon in field theory: a non
renormalizable theory becomes asymptotically free and in fact analytic
in the parameter $\e$ measuring the strength of the perturbation,
because of special symmetries which forbid the exponential growth of
the relevant couplings in absence of free parameters in the Lagrangian
of the model which could possibly be used to control them.
\*

Concerning the originality of the results obtained with the techniques
exposed in this paper the following comments may give an idea of the
status of the matter:
\\
(1) The KAM theory presented above is only a reinterpretation of the
original proofs [El96], following [Ga94] and [GM96a] see also
[BGK98]. However even in classical mechanics the method has generated
new results.  Other problems of the same type that can be naturally
interpreted in terms of renormalization group analysis of suitable
quantum fields, see [GM96a], [GM96b], and in the theory of
the {\it separatrix splitting}, see [Ga95], [GGM99].
\\
(2) The results on the theory of the Fermi systems were obtained for
the first time by the method described above (including the vanishing
of the beta function, [BG90]) and have led to the understanding of
several other problems [BM95], [BGM99], [BM99], [Ma97], [Ma98a],
[Ma98b], [Ma99].

\*\*
{\it Acknowledgements: I am grateful to the Organizing committee of
the conference ``Renormalization Group 2000'' for giving me the
opportunity to write and present this review, and for travel support
to the conference held in Taxco, Mexico. I am indebted to G. Gentile
and V. Mastropietro for precious comments on the draft of this
paper.}
\*\*
%\1
\0{\bf References}
\*
\def\0{\vskip3pt}
\0[BG90] Benfatto, G., Gallavotti G.: {\it Perturbation theory of the Fermi
  surface in a quantum liquid. A general quasi particle formalism and
  one dimensional systems}, Journal of Statistical Physics, {\bf 59},
  541--664, 1990.

\0[BG95] Benfatto, G., Gallavotti, G.: {\it Renormalization
  group}, p. 1--144, Princeton University Press, 1995.

\0[BG99] Berretti, A., Gentile, G.: {\it Scaling properties of the
radius of convergence of a Lindstedt series: the standard map},
Journal de Math\'ematiques, {\bf 78}, 159--176, 1999. And {\it Scaling
properties of the radius of convergence of a Lindstedt series:
generalized standard map}, preprint, 1999.

\0[BGPS] Gallavotti, G., Procacci, A., Scoppola, B.: {\it Beta function
and Schwinger functions for a many body system in one dimension. Anomaly
of the Fermi surface.}, Communications in Mathematical Physics, {\bf
160}, 93--172, 1994.

\0[BGK99] Bricmont J., Gawedzki K., Kupiainen A.: {\it KAM Theorem and
     Quantum Field Theory}, Communications in Mathematical Physics,
     {\bf 201}, 699--727, 1999.

\0[BGM99] Benfatto, G., Gentile, G., Mastropietro, V.:
{\it Electrons in a lattice with incommensurate potential} {\bf
Journal of Statistical Physics}, {\bf 89}, 655-708 (1997)

\0[BM00] Benfatto, G., Mastropietro, V.: {\it A renormalization group
     computation of the spin correlation functions in the $XYZ$
     model}, p. 1--77, preprint, U. Roma 2, january 2000.

\0[BM95] Bonetto, F., Mastropietro, V.:  {\it Beta Function
and anomaly of the Fermi surface for a $d=1$ system of interacting
fermions in a periodic potential}, Communications in Mathematical
Physics, {\bf 172}, 57--93, 1995. See also {\it Filled band Fermi
systems}, {\bf Mathematical Physics Electronic Journal}, {\bf 2},
1--43, 1996. And {\it Critical indices in a $d=1$ filled band Fermi
system}, Physical Review B, {\bf 56}, 1296--1308, 1997.

\0[BGGM] Bonetto, F., Gallavotti, G., Gentile, G., Mastropietro, V.: {\it
Quasi linear flows on tori: regularity of their linearization},
Communications in Mathematical Physics, {\bf192}, 707--736, 1998. And
{\it Lindstedt series, ultraviolet divergences and Moser's theorem},
Annali della Scuola Normale Superiore di Pisa, {\bf 26}, 545--593,
1998, A review in: Gallavotti, G.: {\it Methods in the theory of quasi
periodic motions}, ed. R. Spigler, S. Venakides, Proceedings of
Symposia in Applied Mathematics, {\bf 54}, 163--174, 1997, American
Mathematical Society.

\0[El96] Eliasson, H.: {\it Absolutely convergent series expansions
for quasi-periodic motions}, Ma\-the\-ma\-ti\-cal Physics Electronic
Journal, {\bf 2}, 1996.

\0[Ga95] Gallavotti, G.: {\it Invariant tori: a field theoretic point of
  view on Eliasson's work}, in {\sl Advances in Dynamical Systems and
  Quantum Physics}, 117--132, Ed. R. Figari, World Scientific, 1995.

\0[Ga94] Gallavotti, G.: {\it Twistless KAM tori}, Communications in
Mathematical Physics, {\bf 164}, 145--156, 1994. And Gallavotti, G.,
Gentile, G.: {\it Majorant series convergence for twistless KAM tori},
Ergodic theory and dynamical systems, {\bf 15}, p.  857--869, 1995.

\0[GGM95] Gallavotti, G., Gentile, G., Mastropietro, V.:
  {\it Field theory and KAM tori}, p. 1--9, Mathematical Physics
  Electronic Journal, MPEJ, {\bf 1}, 1995 (http:// mpej.unige.ch).

\0[GGM99] Gallavotti, G., Gentile, G., Mastropietro, V.: {\it Separatrix
  splitting for systems with three time scales}, Communications in
  Mathematical Physics, {\bf202}, 197--236, 1999. And {\it Melnikov's
  approximation dominance. Some examples.}, Reviews on Mathematical
  Physics, {\bf11}, 451--461, 1999.

\0[Ga85] Gallavotti, G.: {\it
  Renormalization theory and ultraviolet stability via
  re\-nor\-ma\-li\-za\-tion
  group methods}, Reviews of Modern Physics {\bf 57}, 471--569, 1985.

\0[Ga94b] Gallavotti, G.: {\it Twistless KAM tori,  quasi flat homoclinic
  intersections, and other
  cancellations in the perturbation series of certain completely
  integrable hamiltonian systems. A review},
  Reviews on Mathematical Physics, {\bf 6}, 343--411, 1994.

\0[GM96a] Gentile, G., Mastropietro, V.: {\it KAM theorem revisited},
Physica D {\bf 90}, 225--234, 1996. And: {\it Tree expansion and
multiscale analysis for KAM tori}, Nonlinearity {\bf 8}, 1159--1178
1995.

\0[GM96b] Gentile, G., Mastropietro, V.: {\it Methods for the
analysis of the Lindstedt series for KAM tori and renormalizability in
classical mechanics. A review with some applications}, Reviews in
Mathematical Physics {\bf 8}, 393--444, 1996.

\0[GS93], Gentile, G., Scoppola, B.: {\it Renormalization group and
the ultraviolet problem in the Luttinger model}, {\bf 154}, 135--179,
1993.

\0[Le87] Lesniewski, A.: {\it Effective action for the Yukawa 2 quantum
field Theory}, Communications in Mathematical Physics, {\bf 108},
437-467, 1987.

\0[LW60] Luttinger, J., and Ward, J. {\it Ground state energy of a many
  fermion system}, Physical Review {\bf 118}, 1417--1427, 1960.

\0[LM66] Lieb, E., and Mattis, D.{\it Mathematical Physics in One
  Dimension}, Academic Press, New York, 1966.  Mattis, D., and
  Lieb, E. {\it Exact solution of a many fermions system and its
  associated boson field}, Journal of Mathematical Physics {\bf 6},
  304--312, 1965. Reprinted in [LM66].

\0[Lu63] Luttinger, J.: {\it An exactly soluble model of a many fermion
  system}, Journal of Mathematical Physics {\bf 4}, 1154--1162, 1963.

\0[Ma97] Mastropietro, V.: {\it Small denominators and anomalous behaviour
in the uncommensurate Hubbard-Holstein model}, mp$\_$arc \#97-652.

\0[Ma98a] Mastropietro, V.: {\it Renormalization group for the XYZ model},
FM 98-13, http:// ipparco. roma1.infn.it.

\0[Ma98b] V. Mastropietro: {\it Renormalization group for the Holstein
Hubbard model}, FM 98-12, http://ipparco.roma1.infn.it.

\0[Ma99] V. Mastropietro: {\it Anomalous BCS equation for a Luttinger
superconductor}, FM 99-1, http://ipparco.roma1.infn.it.

\0[Po92] Poincar\`e, H.: {\it Les M\'ethodes nouvelles de la m\'ecanique
c\'eleste}, 1892, reprinted by Blanchard, Paris, 1987.

\0[P\"o] P\"oschel, J.: {\it Invariant manifolds of complex analytic
mappings}, Les Houches, XLIII (1984), Vol. II, p. 949-- 964, Eds. K.
Osterwalder \& R. Stora, North Holland, 1986.

\0[Pl92]  Polchinski, J.: {\it Effective field theory and the Fermi
  surface}, University of Texas, prep\-rint UTTC-20-92.

\0[Pl84] Polchinski, J.: {\it Renormalization group and effective
  lagr\-angi\-ans}, Nuclear Physics {\bf B231}, 269--295, 1984.

\0[Th83] Thirring, W.: {\it Course in Mathematical Physics}, vol. 1,
p. 133, Springer, Wien, 1983

\0[El96] Eliasson, L.H.: {\it Absolutely convergent series expansions for
quasi-periodic motions}, Mathematical Physics Electronic Journal
{\bf 2}, 1996.

\0[So79] Solyom, J.: {\it The Fermi gas model of one dimensional
conductors}, Adv. Phys., {\bf 28}, 201-303, (1979).

\0[Wi70] Wilson, K. G. {\it Model of coupling constant
  renormalization}, Physical Review {\bf D2}, 1438--1472, 1970.

\0[WF72] Wilson, K. G., and Fisher, M. {\it Critical exponents in $3.99$
  dimensions}, Physical Review Letters {\bf 28}, 240--243, 1972.

\0[Wi73] Wilson, K. G. {\it Quantum field theory models in less than four
  dimensions}, Physical Review {\bf D7}, 2911--2926, 1973

\0[WK74] Wilson, K. G., and Kogut, J. B. {\it The renormalization group
  and the $\e$-expansion}, Physics Reports {\bf 12}, 76--199, 1974.

\let\0=\noindent
\FINE

\ciao